\documentstyle[psfig,12pt,a4]{article}
\begin{document}
\thispagestyle{empty}
\rightline{HUB-EP-97/21}
\rightline{LMU-HEP-97-5}
\rightline{SFB-375/164}
\rightline{hep-th/9704013}
\vspace{1truecm}

\centerline{\bf \Large 
Instanton effects in string cosmology}

\vspace{1.2truecm}
\centerline{\bf Klaus Behrndt$^a$,
Stefan F\"orste$^b$
and Stefan Schwager$^b$\footnote{e-mail: \tt 
behrndt@qft2.physik.hu-berlin.de, 
Stefan.Foerste@physik.uni-muenchen.de, \hfill \\ 
Stefan.Schwager@physik.uni-muenchen.de }}
\vspace{.5truecm}
{\em
\centerline{$^a$ Humboldt-Universit\"at, Institut f\"ur Physik}
\centerline{Invalidenstra\ss e 110, 10115 Berlin, Germany}
\vspace{.3truecm}
\centerline{$^b$ Sektion Physik, Universit\"at M\"unchen}
\centerline{Theresienstra\ss{}e 37, 80333 M\"unchen, Germany}
}

\vspace{2.2truecm}


\vspace{.5truecm}

\begin{abstract}
We consider the gauge dyonic string solution of the $K3$ compactified
heterotic string theory in a four dimensional cosmological context.
Since for this solution Green-Schwarz as well as Chern-Simons
corrections have been taken into account it contains both
world sheet and string loop corrections. The cosmological
picture is obtained by rotating the world volume of the gauge
dyonic string into two space like dimensions and compactifying
those dimensions on a two torus. We compare the result
with gauge neutral extreme and non-extreme cosmologies
and find that the non-trivial Yang Mills background leads to
a solution without any singularities whereas for trivial Yang-Mills
backgrounds some of the fields become always singular at the
big bang.
\end{abstract}

\let\LARGE=\large
\let\Large=\large
\let\large=\normalsize

\vfill
{\small

}

\newpage

\section{Introduction}
Since string theory is supposed to describe physics at scales
which are not reachable in any terrestrial laboratory cosmological
models predicted by string theory might be the suitable 
candidates for facing this theory with nature.  Therefore there has been
quite some interest in studying cosmological vacua of string theory
[\ref{first}-\ref{last}]. A major approach in studying string cosmology
is based on the pre-big-bang scenario [\ref{ven1}-\ref{venlast}],
(for a recent review containing additional references on the topic
see \cite{venlast}). In the pre-big-bang scenario two
cosmological solutions related by scale factor duality
are connected
in a strong coupling region. Since one of the solution describes an
accelerated expanding universe (inflationary branch) whereas the
second solution corresponds to a decelerated expanding universe
(FRW branch) this scenario has many realistic phenomenological
implications. However, it suffers from some open questions.
Because the strong coupling region is hardly accessible
one outstanding problem is the description of the details
how the two solutions are smoothly connected (graceful
exit problem) \cite{brustein}. 

\medskip

\noindent
The second track one can follow is to find more general 
cosmological string vacua, e.g.\  with non vanishing spatial
curvature. Here, one can employ the 
considerable progress in the understanding
of low-energy solutions of string theory which has been made
during the last years. Hopefully, this approach will finally converge
with the one described above. 
A better understanding of low-energy solutions of string theory
was achieved by interpreting
those solutions as intersections of the $D$-brane solutions of type II
string theory. First one reinterpreted the known 4-d black
holes as bound states of brane solution. But the procedure is
quite general and can also be used to find new solutions in 4
dimensions, not only of 0-brane type (black holes) but also of 1-brane
(string) or (-1)-brane type (instanton). For a classification of all
multiple intersections of $D$-branes see \cite{eric}.

\medskip

\noindent
In the present paper we are going to discuss 4-d cosmological solutions 
obtained
by  compactification of 6-d heterotic string solutions. 
The standard cosmology is
described in terms of a Robertson-Walker metric
\begin{equation}
 ds^2 = -d\tau^2 + K^2(\tau) d\Omega_{3,k} 
\end{equation}
where $K$ is the world radius and $d\Omega_{3,k}^2 = d\chi^2 +
\sin^2(\sqrt{k} \chi) / k(d\theta^2 + \sin^2\theta d\omega)$,
which is the 3-d spherical volume measure with the curvature $k= -1,0,1$.
This metric ansatz takes into account that the universe is
homogeneous and isotropic (as a good approximation). 
Translated into the brane picture this means, that the Robertson-Walker
space-time describes a (-1)-brane, i.e.\ the big bang is a point in the
4-d space-time. 

\smallskip

Following this philosophy, there are two ways to obtain cosmological
solutions from intersecting branes. The first one is to consider already
a (-1)-brane in 10 dimensions and as second way one can consider the
``standard'' brane intersection and wraps the complete world volumes of
the branes into the internal space.  In order to be able to wrap
the world volume of the brane one has to rotate the world volume
such that the time coordinate is orthogonal to it.  This can be done by
performing a  special analytic continuation (Wick-rotation) or by taking a 
non-extreme brane solution and
going behind the outer horizon where the time and radius coordinates
interchange. 

\medskip

\noindent
In heterotic
string theory, one has the fundamental string and $NS$-5-brane 
solution in 10-d. Wrapping the 5-brane world volume into the internal 
space one gets
immediately a cosmological solution in 4-d. Making this solution
non-extremal and going behind the horizon one obtains a cosmological
model that includes  all possible values for the spatial curvature $k$
\cite{behrndt, wands}. Since the $NS$-5-brane is part of all superstring
theories one can explore the type IIB $S$-duality in order to convert the
brane into a $RR$-5-brane \cite{schwager-poppe}. This way one can
get some insight into strong coupling corrections to
cosmological string vacua since $RR$ charged states are non
perturbative states. A more systematic
approach to cosmological solutions coming from any type of branes 
has been discussed in \cite{ovrut}. The field equations 
of any 
(non-extreme) brane solution, which shall be wrapped completely into 
the internal space can be reduced to 1-d field equations. In special 
cases one can solve these equations or use known solutions, 
e.g.\ in terms of Toda models \cite{kaloper}. In this procedure one has, 
however, to take into account that the intersections of extreme branes 
always yield a trivial Einstein frame metric \cite{tseytlin-cosmo}. Since
in these models the big bang appears as the surface of the $D$-branes
one can argue, that these branes could yield to a resolution of
the big bang singularity \cite{larsen}.

\medskip

\noindent
In the present paper we will find that the inclusion of world sheet
loop corrections ($\alpha^\prime$) in addition to string loop
corrections ($e^\phi$) might lead to a completely non singular
universe. A solution containing the two kinds of corrections
is the gauge dyonic string constructed in \cite{dlp}. This solution
contains an $SU(2)$ instanton and thus Chern-Simons and 
Green-Schwarz corrections are non-trivial.
Our aim in this paper is to investigate those ``instanton effects''
in a four dimensional cosmological context.
Therefore, our starting point is a heterotic 6-d model, which corresponds
in 10-d to a fundamental string lying inside an $NS$-5-brane (see figure
1). We will compactify this heterotic solution on a $K3$ manifold
yielding an $N=2$ string cosmology in 4 dimensions after further
compactification on $T^2$.

\smallskip

The paper is organized as follows. In the next section we will
discuss the $K3$ compactified heterotic string and describe the
toroidal compactification down to four dimensions. 
In section 3 we study various six dimensional vacua describing
(macroscopic) strings. By an analytic continuation we rotate the
world volume of those strings into space-like directions and obtain
the (-1)-brane configurations required for a cosmological
interpretation. 
Section 4 deals with  the resulting four dimensional cosmologies. We will
first review known results for the instanton-free case and afterwards we
will discuss the special effect of the YM instantons. We will
also comment on the corresponding brane picture. Finally, we will
summarize our results. 

\begin{figure}[t]
\begin{center}
\begin{minipage}{8cm}
\centerline{}
\psfig{figure=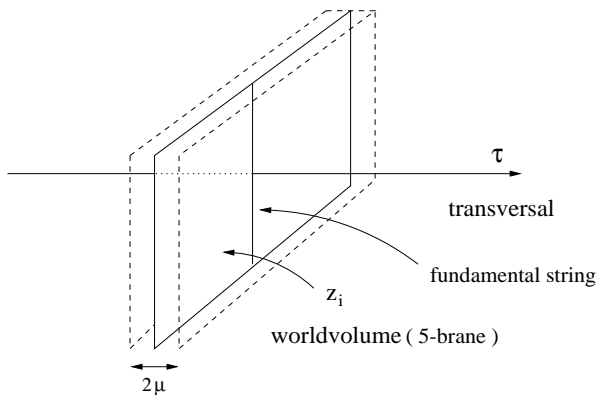,width=8cm}
\end{minipage}\\
\end{center}
\vspace{-6mm}
\caption{
This figure shows a fundamental string lying in a $NS$-5-brane.
The dashed line indicates the non-extreme version of this
intersection. Note, there is a horizon only for $k=+1$. 
We have rotated this configuration so that the time $\tau$ is
the transverse direction.
}
\end{figure}
\section{The heterotic string theory on $K3\times T^2$} \label{reduction}
In this section we will develop the general setting for compactifying
the $K3$ compactified heterotic string theory further down on $T^2$.
The bosonic part of the six dimensional vacuum satisfies the following
set of equations \cite{sagnotti} ($M, N = 0, \ldots ,5$)
\begin{equation} \label{dilaton-eq}
\Box \phi = -\frac{1}{12} e^{-2\phi} H^2 + \frac{\alpha^\prime}{16}
\sum_{\alpha} \left( v_{\alpha} e^{-\phi} - \tilde{v}_{\alpha} e^{\phi}
\right) tr  F_{\alpha}^2 ,
\end{equation}
\begin{equation} \label{einstein} \begin{array}{ll}
R_{MN} = & \partial_M\phi\partial_N\phi +\frac{1}{4} e^{-2\phi } \left(
H^2 _{MN} - \frac{1}{6} H^2 G_{MN}\right) \\
 & -\frac{\alpha^\prime }{4} \sum_{\alpha} \left( v_{\alpha} e^{-\phi}
+ \tilde{v}_{\alpha} e^{\phi}\right) tr  \left( F_{\alpha\, MN}^2 
-\frac{1}{8} F_\alpha ^2 G_{MN}\right) ,
\end{array} \end{equation}
\begin{equation} \label{yang-mills}
D_M \left( \left( v_{\alpha} e^{-\phi}+ \tilde{v}_{\alpha} e^{\phi}\right) 
F_{\alpha}^{MN} \right) -\frac{1}{2} v_{\alpha} e^{-2\phi} H^{N}_{PQ} 
F_{\alpha}^{PQ} 
-\frac{1}{2}\tilde{v}_{\alpha}\mbox{} ^*H^{N}_{PQ}F^{PQ}_{\alpha}=
0 ,
\end{equation}
\begin{equation} \label{bianchi}
dH = \frac{\alpha^\prime}{4} \sum_{\alpha} v_{\alpha} tr F_{\alpha}\wedge
F_{\alpha},
\end{equation}
\begin{equation} \label{H-field-eq}
d\tilde{H} = \frac{\alpha^{\prime}}{4} \sum_{\alpha} 
\tilde{v}_{\alpha} tr F_{\alpha} \wedge F_{\alpha} .
\end{equation}
where $\tilde{H} = e^{-2\phi}\mbox{}^* H$.
The solution we are going to consider has vanishing 
Lorentz Chern-Simons forms and vanishing $R\wedge R$ and we
did not include those contributions into the above set of equations.
Terms with $v_{\alpha} F_{\alpha} \wedge F_{\alpha}$ are due to
gauge Chern-Simons corrections whereas terms with $\tilde{v}_\alpha
F_\alpha \wedge F_\alpha $ originate from the Green-Schwarz term.
(Here, $\alpha $ labels an unbroken non Abelian subgroup 
of $E_8\times E_8$
or $SO(32)$ and we assume that the unbroken gauge symmetry is a direct
product of non Abelian factors, only.) 
The values for $v$ and $\tilde{v}$ depend on the instanton embedding
in the internal dimensions and can be 
found e.g.\ in appendix B of \cite{dmw}.
For the symmetric embedding of 12 instantons in each $E_8$ of 
the $E_8\times E_8$ string $v$ is always positive whereas the perturbative
value for $\tilde{v}$ vanishes. It has been conjectured however that
$\tilde{v}$ becomes equal to $v$ due to non perturbative effects
\cite{dmw, berkooz}. For the asymmetric instanton embedding and the
$SO(32)$ string $\tilde{v}$ is negative and a phase transition occurs 
when the gauge kinetic term changes sign \cite{dmw,mv,afiq,dlp}.

The six dimensional vacua we are interested in fit into the following ansatz
 ($\mu , \nu =0, \ldots, 3$),
\begin{equation} \label{6-metric}
ds_6^2 = e^{-A(x)}\hat{g}_{\mu\nu}(x)dx^\mu dx^\nu +
\frac{e^{A(x)}}{U_2}\left( dy + U dz\right) \left( dy +\bar{U}dz\right) ,
\end{equation}  
where 
\begin{equation}
U = U(x) = U_1 + i U_2
\end{equation}
is the complex structure of the torus we are going to compactify on.
For the fields we choose the following ansatz, 
\begin{equation}
\phi = \phi \left( x\right)
\end{equation}
\begin{equation} \label{eq:b}
H_{yz\mu} =   \partial_{\mu} b(x),
\end{equation} 
\begin{equation}\label{eq:axion}
H_{\mu\nu\rho} = e^{2\phi -2A}\sqrt{\hat{g}}\epsilon_{\mu\nu\rho\lambda} 
\hat{g}^{\lambda\kappa}\partial_{\kappa} a 
\end{equation}
\begin{equation}
F^{\alpha}_{\mu\nu} = F^{\alpha}_{\mu\nu} \left( x\right)
\end{equation}
and the rest of the fields is zero.

Now, we are going to perform the dimensional reduction on a two torus
with the coordinates $y$, $z$ at the level of the action. For simplicity
we neglect Green-Schwarz terms ($B\wedge F_\alpha\wedge F_\alpha$)
at the moment and deduce their
contribution in the end by general arguments. 
Since $F_{yz} =F_{y\nu}=F_{z\nu}=0$
those terms will give 
rise to $ b \tilde{v}_\alpha F_\alpha \wedge F_\alpha$ contributions to the
four dimensional Lagrangian. Without Green-Schwarz terms the
six dimensional action is \footnote{ This is for the convention 
that the universal sector of the effective string frame action is given by
$ S \sim \int \sqrt{G}e^{-2\phi}\left[ R + 4\left( \partial \phi\right)^2 
-\frac{1}{12} H^2\right] $.}
\begin{equation} \label{sag-action}
S_6 \sim \int d^6x \sqrt{G_6}\left\{ R_6 - \left( \partial \phi\right)^2
-\frac{e^{-2\phi}}{12} H^2 -\frac{\alpha^\prime}{8}
\sum_\alpha \left( v_\alpha e^{-\phi}
+\tilde{v}_\alpha e^{\phi}\right) tr F_\alpha ^2\right\},
\end{equation}
The ansatz (\ref{eq:axion}) can be thought of as just using 
$H_{\mu\nu\rho} = H_{\mu\nu\rho}\left( x\right)$ and going to the dual
axion $a$ in four dimensions. This is done by adding the Bianchi
identity (\ref{bianchi}) to the action with a Lagrange multiplier $a$
and integrating out $H$. The Chern-Simons corrections will result
in terms containing $a v_\alpha F_\alpha \wedge F_\alpha $. At the moment
we drop also these contributions and reinstall them in the end together
with the above neglected Green-Schwarz terms.
The form of the metric (\ref{6-metric})
gives rise to the following expression for the scalar curvature, (scalar
products, covariant derivatives  are taken with respect to $\hat{g}$),
\begin{equation}
R = e^{A}\left\{ \hat{R}+\hat{\nabla} ^2 A +2 \frac{\partial U \partial \bar{U}}{\left(
U - \bar{U}\right)^2}  -\left( \partial A\right)^2 \right\} .
\end{equation}
In the compactified theory it is convenient to  redefine fields according to
\begin{equation} \label{varphi}
A =  \lambda -\varphi,
\end{equation}
\begin{equation} \label{lambda}
\phi = \varphi + \lambda
\end{equation}
and to define the complex axion-dilaton field $S$ as
\begin{equation}
S= S_1 + i S_2= a + i e^{-2\varphi}
\end{equation}
and the complex K\"ahler structure $T$ as
\begin{equation}
T = T_1 + i T_2 = b + i e^{2\lambda}
\end{equation}
Integrating (\ref{sag-action}) over the two torus (labeled by $y$, $z$) results in 
\begin{equation}
\begin{array}{ll}
S \sim & \int d^4x\sqrt{\hat{g}}\left\{ \hat{R} 
+ 2   \frac{\partial U \partial \bar{U}}{\left(
U - \bar{U}\right)^2} + 2   \frac{\partial S \partial \bar{S}}{\left(
S - \bar{S}\right)^2}+ 2   \frac{\partial T \partial \bar{T}}{\left(
T - \bar{T}\right)^2}  \right. \\
&\left. -\frac{\alpha^\prime}{8} \sum_{\alpha}\left( v_\alpha S_2+
\tilde{v}_\alpha T_2
\right) tr F_{\alpha}^2\right\} .\end{array}
\end{equation}
Now, let us infer the Chern-Simons and Green-Schwarz corrections
by using general arguments. The gravitational Green-Schwarz and
Chern-Simons correction are of higher order in the derivatives
and can be neglected at the considered order. In addition, for the
vacuum we are going to study $R\wedge R$ vanishes identically.
As discussed before the gauge field dependence of the neglected
contributions will be of the form
$tr F\tilde{F} $ with $\tilde{F}$ being the Hodge dual of $F$ (in four
dimensions). Supersymmetry requires holomorphic gauge couplings and
that fixes the final expression of the reduced action to \cite{wb}
\begin{equation}
\begin{array}{ll}
S \sim & \int d^4x\sqrt{\hat{g}}\left\{ \hat{R} 
+ 2   \frac{\partial U \partial \bar{U}}{\left(
U - \bar{U}\right)^2} + 2   \frac{\partial S \partial \bar{S}}{\left(
S - \bar{S}\right)^2}+ 2   \frac{\partial T \partial \bar{T}}{\left(
T - \bar{T}\right)^2}  \right. \\
&\left. - \frac{\alpha^\prime}{8}
\sum_{\alpha} Im\left[ \left( v_\alpha S+\tilde{v}_\alpha T
\right) \left( tr F_{\alpha}^2
-\frac{i}{2} tr F_{\alpha}\tilde{F}_{\alpha}\right)\right] \right\} .\end{array}
\end{equation}
These gauge field couplings follow also from an holomorphic 
prepotential as described e.g.\ in \cite{gabriel}.
\section{The six dimensional solution}
In this section we will discuss the six dimensional vacuum which will
lead us to a four dimensional cosmological solution upon toroidal
compactification. First, we will discuss a string solution with a trivial
Yang-Mills background and the analytic continuation 
giving the four dimensional cosmological interpretation. This will
be done for the extreme and the non-extreme solution. In the extreme
case we can switch on a non-trivial Yang-Mills background.
That will be a modified form of the instanton
solution discussed in \cite{dlp}. The unbroken gauge symmetry in 
six dimensions is taken to be $SU(2)$. 
After an analytic continuation the
gauge field will take values in the subset of $SL(2,C)$ that can be viewed
as a Euclideanized version of $SL\left(2,I\!\! R\right)$.
\subsection{Extreme and non-extreme vacuum with vanishing Yang-Mills
background}
The non-extreme dyonic string solution has been constructed in
\cite{dulupo} starting from the extreme solution of \cite{dfkr}. The issue how 
to obtain non-extreme solutions from extreme ones is also discussed in
\cite{cvetic}.  For completeness we give the dyonic black string
solution in six dimensions before performing the analytic continuation.
\begin{equation}
ds^2 = e^A \left( -dt^2 e^{2f} + dz^2\right) + e^{-A}\left( dr^2 e^{-2f}
+r^2 d\Omega_3 ^2 \right) ,
\end{equation}
with
\begin{equation}
d\Omega_3 ^2 = d\chi^2 + \sin^2\!\chi \left( d\theta ^2 +\sin^2\!\theta 
d\omega^2\right) ,
\end{equation}
\begin{equation}
e^{2f} = 1 - \frac{\mu^2}{r^2} .
\end{equation}
For the rest of the fields one finds
\begin{equation} \label{phi1}\begin{array}{ll}
e^{2\varphi} & = 1 + \frac{\mu^2}{r^2} \sinh^2\!\alpha \\
e^{-2\lambda} & = 1+ \frac{\mu^2}{r^2} \sinh^2\! \beta ,
\end{array}
\end{equation}
with $\varphi$ and $\lambda$ as in (\ref{varphi}) and (\ref{lambda}).
The $H$ field is the sum of a contribution coupling to the electric
(fundamental) string and a contribution coupling to the
magnetic string,
\begin{equation}
H = H_e + H_m ,
\end{equation}
with
\begin{equation}\label{H1}\begin{array}{ll}
H_e & = \coth\! \beta\, d e^{2\lambda}\wedge dt \wedge dz ,\\
H_m &=  \coth\! \alpha\, r^3\partial_r e^{2\varphi}
\sin^2\! \chi \sin\theta d\chi \wedge d\theta \wedge d\omega.
\end{array}\end{equation}
The extreme solution is obtained by performing the limit
$\mu \rightarrow 0$, $\alpha \rightarrow \infty$ and 
$\beta \rightarrow \infty$ such that
\begin{equation}
\mu^2 \sinh^2\!\alpha \rightarrow P \,\,\, , \,\,\,
\mu^2 \sinh^2\! \beta \rightarrow Q,
\end{equation}
with $P$ and $Q$ finite.
In our context another branch of the solution will be important. 
That branch
is not considered in \cite{dulupo} since fields become singular at 
 $r^2 = \mu^2\cosh^2\!\alpha$ and $r^2=\mu^2\cosh^2\!\beta$. 
In the second solution one finds
\begin{equation}\label{phi2} \begin{array}{ll}
e^{2\varphi} & = 1 - \frac{\mu^2}{r^2} \cosh^2\!\alpha \\
e^{-2\lambda} & = 1- \frac{\mu^2}{r^2} \cosh^2\! \beta ,
\end{array}
\end{equation}
and
 \begin{equation}\label{H2}\begin{array}{ll}
H_e & = \tanh\! \beta\, d e^{2\lambda}\wedge dt \wedge dz ,\\
H_m & =  \tanh\! \alpha\, r^3\partial_r e^{2\varphi}
\sin^2\! \chi \sin\theta d\chi \wedge d\theta \wedge d\omega.
\end{array}\end{equation}
In the non-extreme case there are basically two options to obtain
a solution serving our purpose of obtaining a four dimensional
cosmological background. In order to interpret $t$ and $z$ as 
coordinates of an internal torus one needs to change the sign
of the $tt$ component of the metric. In addition, a time-like 
coordinate in the compactified solution is required. A homogeneous
and isotropic 4-d-time dependent solution is obtained by flipping
the signature of the $rr$ metric component. For the non-extreme
solution this happens when we take $r$ to be in the region
between the inner and the outer horizon, i.e.\ $r^2 < \mu^2$.
The second option which we will consider in the rest of the paper 
is given by performing a Wick rotation
\begin{equation}
t=iy,
\end{equation}
\begin{equation}
r = i\tau ,
\end{equation}
\begin{equation}
\chi \rightarrow i\chi .
\end{equation}
Note that the Wick rotation here and in the following
does not change the signature of the 6-d theorie,
it remains a Minkowskian theory. Therefore, the
action as well as the equations of motion remain
the same and the rotated solutions still solve
eqs.\ (\ref{dilaton-eq}) - (\ref{H-field-eq}), however, with a non-compact
gauge group (see next section). The aim of this
rotation is to convert the radius
into a timlike direction and the time into a spatial
direction.
The continued solution reads
\begin{equation}
ds^2 = e^A\left( dy^2 e^{2f} + dz^2\right)+ e^{-A} \left(-d\tau^2 e^{-2f}
+ \tau^2 d\Omega_{3,-1}^2 \right) ,
\end{equation}
\begin{equation}
d\Omega_{3,-1}^2 = d\chi^2 + \sinh^2\!\chi\left( d\theta^2
+ \sin^2\!\theta d\omega^2\right),
\end{equation}
\begin{equation}
e^{2f} = 1+\frac{\mu^2}{\tau^2} ,
\end{equation}
\begin{equation} \label{cphi1}\begin{array}{ll}
e^{2\varphi} & = 1- \frac{\mu^2}{\tau^2} \sinh^2\! \alpha ,\\
e^{-2\lambda} & = 1 - \frac{\mu^2}{\tau^2} \sinh^2\! \beta,
\end{array} \end{equation}
and
\begin{equation} \label{imi}\begin{array}{ll}
H_e =& i\coth \!\beta\, de^{2\lambda}\wedge dy \wedge dz  ,\\
H_m =& i\coth \!\alpha\, \tau^3\partial_\tau e^{2\varphi} 
\sinh^2\!\chi \sin\theta d\chi 
\wedge d\theta
\wedge d\omega  ,\end{array}
\end{equation}
for the first branch (\ref{phi1}), (\ref{H1}). Whereas the second branch
(\ref{phi2}), (\ref{H2}) continues to
\begin{equation} \label{cphi2} \begin{array}{ll}
e^{2\varphi} & = 1+ \frac{\mu^2}{\tau^2} \cosh^2\! \alpha ,\\
e^{-2\lambda} & = 1 + \frac{\mu^2}{\tau^2} \cosh^2\! \beta,
\end{array} \end{equation}
and
\begin{equation} \label{imag}\begin{array}{ll}
H_e =& i\tanh\!\beta\, de^{2\lambda}\wedge dy \wedge dz ,\\
H_m =& i\tanh\!\alpha\, \tau^3\partial_\tau e^{2\varphi} \sinh^2\!\chi 
\sin\theta d\chi 
\wedge d\theta
\wedge d\omega .\end{array}
\end{equation}
Taking in addition now $\alpha$, $\beta$ to $i\alpha$, $i\beta$ gives
real magnetic and electric charges. In that case 
both solutions are related by replacing $\alpha$, $\beta$
with $\frac{\pi}{2} -\alpha$, $\frac{\pi}{2} -\beta$ and the only 
singularity ist at $\tau =0$. However, we will be interested in
performing the extreme limit and have therefore to stick to real
$\alpha$ and $\beta$. In order to have a singularity only
at $\tau=0$ (the position of the brane)\footnote{This requirement
corresponds to our picture of rotating the world volume of the brane
which should not lead to additional singularities.} we have to take 
solution (\ref{cphi2}),
(\ref{imag}) 
and encounter the problem of having imaginary charges in the
continued solution. We will give a conjectural  resolution of that
problem in paragraph \ref{probe-string}.
\subsection{Extreme vacuum with non-trivial Yang-Mills
background \& analytic continuation}
Now, we will analytically continue the gauge dyonic string
solution of \cite{dlp}. For that purpose let us rederive their solution
in a way that allows us to perform 
a ``continuous Wick rotation''.
That means that we take as a metric ansatz
\begin{equation}
ds_6^2 = e^A\left( -qdt^2+dz^2\right) + e^{-A}\left( \frac{dr^2}{q}
+r^2 d\Omega^2 _{3,k} \right),
\end{equation}
with the $S_{k}^3$ measure
\begin{equation}
d\Omega_{3,k}^2 = d\chi ^2 +\frac{\sin^2\! \left( \sqrt{k} \chi\right)}{k}
\left( d\theta ^2 + \sin^2 \!\theta d\omega^2\right) ,
\end{equation}
and $q,k$ are real constants. A ``continuous Wick rotation'' is performed
by taking $q$ from positive values through zero to negative values.
The constant $k$ will be fixed by the equations of motion. 
We take the unbroken gauge symmetry to be $SU(2)$ and 
the gauge field strength to be self dual with respect to
\begin{equation}
ds_4 ^2 =  \frac{dr^2}{q} + r^2 d\Omega_{3,k}^2 .
\end{equation}
An $S_{k}^3$ symmetric ansatz for the gauge field is given by
\begin{equation} \label{grupp}
{\cal A}\left( r\right) = \gamma\left( r\right) \sqrt{q}g^{-1} dg,
\end{equation}
with
\begin{equation} \begin{array}{ll}
g = &  \cos \sqrt{k} \chi - i \cos \omega \sin \theta \sin \sqrt{k} \chi 
\sigma_1 - i \sin \omega \sin \theta \sin \sqrt{k} \chi \sigma_2\\ &
-i \cos \theta \sin \sqrt{k} \chi \sigma_3 \, . \end{array}
\end{equation}
The self duality condition leads to the following differential equation
\begin{equation}
r\gamma^\prime = 2\gamma \left(1-\sqrt{q}\gamma\right)\sqrt{\frac{q}{k}}
\end{equation}
with the solution
 \begin{equation}
\gamma\left( r\right) = \frac{\exp \left\{ \sqrt{\frac{q}{k}} 
\log \frac{r^2}{\rho^2 \sqrt{q}}
\right\} }{\sqrt{q}\exp \left\{ \sqrt{\frac{q}{k}}\log \frac{r^2}{\rho^2\sqrt{q}}\right\}
+1} ,
\end{equation}
where $\rho^2 \sqrt{q}$ is an integration constant.
The ansatz for the $H$-field is taken to be
\begin{equation}
H_{\chi\theta\omega} =\sqrt{q} \frac{r^3 \sin^2 \sqrt{k}\chi \sin\theta}{k}
\partial_r e^C ,
\end{equation}
\begin{equation}
H_{ijr} = \sqrt{q}\epsilon_{ij} \partial_r e^{\tilde{C}} .
\end{equation}
The Yang-Mills equation (\ref{yang-mills}) is then satisfied if we choose 
$C=\varphi$ and $\tilde{C}=\lambda$. The $H$-field equation 
(\ref{H-field-eq}) and the Bianchi 
identity (\ref{bianchi}) lead to
\begin{equation} \label{magnetic}
\Box _0 e^{\phi - A} = \frac{e^{-2A}}{8q} v \alpha^\prime tr F^2 ,
\end{equation}
\begin{equation} \label{electric}
\Box _0 e^{-\phi - A} = \frac{e^{-2A}}{8q} \tilde{v} 
\alpha^\prime tr F^2 ,
\end{equation}
where $\Box_0 = \frac{1}{r^3}\partial_r \left(r^3 \partial_r\right)$
is the flat Laplacian.
The rest of the equations of motions is automatically satisfied
apart from the angular components of the Einstein equations 
(\ref{einstein}) which provide the condition
\begin{equation} \label{constraint}
k=q .
\end{equation}
Finally, $ e^{-2A}trF^2$ is given by
\begin{equation}
\frac{e^{-2A}}{4}tr F^2 = -24 q \frac{\rho^4}{\left( r^2 + \rho^2\right)^4} ,
\end{equation}
and with (\ref{magnetic}) and
(\ref{electric}) we have just rederived the equations given 
in \cite{dlp}. After a further numerical rescaling of the group generators
we can copy their solution,\footnote{With (\ref{constraint})
our metric ansatz and the one in \cite{dlp} are connected by simple
coordinate transformations.}
\begin{equation} 
\begin{array}{ll}
e^{2\varphi} = & e^{\phi_0} + \frac{2v\alpha^\prime
\left( 2\rho^2 + \tau^2\right)}{\left( \rho^2 +
\tau^2\right)^2}, \\
e^{-2\lambda} = &  e^{-\phi_0} + \frac{2\tilde{v}\alpha^\prime
\left( 2\rho^2 + \tau^2\right)}{\left(
 \rho^2 +\tau^2\right)^2}.
\end{array} \end{equation}
This rederivation is useful since it
teaches us some interesting facts about the analytic continuation.
At the first sight it seems to be quite disappointing that we obtained the
constraint (\ref{constraint}) since that implies that we will not be able
to get cosmological solutions with a flat or a spherical three-space.
On the other hand this condition is logical for the following reasons.
When we send $q=k$ to zero the spatial $S^3$ decompactifies to a 
flat space and hence the instanton solution loses its topological
meaning of mapping the spatial $S^3$ onto the $SU(2)$-$S^3$.
Considering the Cartan-Killing 
metric $tr \!\left( \sqrt{q} g^{-1}dg\right)^2$
one observes that it remains negative definite also after the
analytic continuation to negative $q$. That is, after the analytic 
continuation
the gauge field takes values in the subset of $SL(2,C)$ that can be
viewed as an Euclideanized version of $SL(2, I\!\! R)$. So, we continue
also the gauge group from one real realization  ($SU(2)$) to another
one ( $SL(2,I\!\! R)_E $). Topologically we continue
from a $S^3$ group manifold to a set of group elements with
$S^3 _{-1}$ topology and for negative values of $q$ the ``instanton''
maps a pseudo-sphere in the target space on a pseudo-sphere in the
group space. In the following we will call this also instanton and drop
the quotation marks.
\subsection{Analytic continuation in the 
string-probe action}\label{probe-string}
In this paragraph we are going to study how the analytic continuation
(rotating the world volume of the string) is seen from the view
point of a probe string. Thereby we will give a possible resolution
of the imaginary charge problem (\ref{imag}).
Here, we will focus on the extreme solutions. Let us recall
the (for the present discussion) relevant parts of the extreme
solution. In the string frame the extreme 6-d string solution is of the
following form
\begin{equation} \label{string}
\begin{array}{ll}
 ds_6 ^2  =& e^{2\lambda}\left( - dt^2 + dz^2\right) + e^{2\varphi} 
dx^\mu dx^\mu ,\\
H_e =& d\left( e^{2\lambda} dt \wedge dz\right) ,
\end{array}
\end{equation} 
whereas the continuation is of the form
 \begin{equation} \label{-1-brane}
\begin{array}{ll}
 ds_6 ^2  =& e^{2\lambda}\left( dy^2 + dz^2\right) + e^{2\varphi} 
\left( - d\tau^2 + \tau^2 d\Omega_{3,-1} ^2 \right) ,\\
H_e =& id\left( e^{2\lambda} dy \wedge dz\right) .
\end{array}
\end{equation}
First, we consider a string-probe in the 6-d string-background. 
The probe action is 
\begin{equation} \label{probe-action1}
S= -\frac{1}{4\pi \alpha^\prime} \int d^2\sigma \sqrt{ \det\left(
- g_{MN}\partial X^M \partial X^N\right)} + 
\frac{1}{4\pi \alpha^\prime}\int B^{(2)},
\end{equation}
where $M,N = 0 , \ldots , 5$ and $H= dB^{(2)}$.
Gauge fixing the world sheet diffeomorphisms via, ($\sigma^i$ are the
Minkowskian world sheet parameters),
\begin{equation}
t = \sigma^0 \,\,\, , \,\,\, z = -\sigma^1 \,\,\, , \,\,\, X^\mu = X^\mu \left( \sigma^0
\right)
\end{equation}
and plugging (\ref{string}) with $B^{(2)} = e^{2\lambda} dt\wedge dz$
into (\ref{probe-action1}) results in
\begin{equation}
S = \frac{R}{8\pi \alpha^\prime}\int d\sigma^0 e^{2\varphi} v^2 + {\cal O}
\left( v^4\right) ,
\end{equation}
with $v^\mu = \frac{dX^\mu}{d\sigma^0}$ being a four dimensional
velocity vector and  $R$ is the length of the string. 
(In difference to \cite{douglas} we have no contribution from the
string at rest since we included an additional constant into $B^{(2)}$.)
The arising picture is that the string-probe is extended along the
$z$-axis and follows geodesics with respect to
\begin{equation}
ds^2 = e^{2\varphi} dX^\mu dX^\mu ,
\end{equation}
with changing time $\sigma^0$.

\medskip

\noindent
In addition to the
analytical continuation in target space we perform also a continuation
in the probe action, namely we rotate the world volume of the probe
such that the time direction is orthogonal to it. This is done by
a Wick-rotation
\begin{equation}
\sigma^0 \rightarrow i \sigma^0 ,
\end{equation}
 modifying the probe-action (\ref{probe-action1}) to
\begin{equation} \label{probe-action2}
S= -i\frac{1}{4\pi \alpha^\prime} \int d^2\sigma \sqrt{ \det\left(
g_{MN}\partial X^M \partial X^N\right)} + 
\frac{1}{4\pi \alpha^\prime}\int B^{(2)}.
\end{equation}
Now, we choose as a gauge 
\begin{equation} \begin{array}{l}
y = \sigma^0 \,\,\, , \,\,\, z= -\sigma^1 \,\,\, , \,\,\, \tau = \tau\left( \sigma^0 
\right) \\
\chi = \chi\left(\sigma^0\right) \,\,\, , \,\,\, \theta = \theta \left( \sigma^0\right)
 \,\,\, ,
\,\,\, \omega = \omega\left( \sigma^0\right)
\end{array}\end{equation}
and plug (\ref{-1-brane}) into (\ref{probe-action2}) with 
$B^{(2)} = ie^{2\lambda} dy\wedge dz$ . The result is
\begin{equation}
S = \frac{R}{8\pi \alpha^\prime}\int d\left( i\sigma^0\right) e^{2\varphi} \left\{ 
\left(\frac{d\tau}{d\sigma^0}\right)^2\!\!\! - \tau^2\! \left[ \left(\frac{d\chi}
{d\sigma^0}\right)^2\!\!\!
+ \sinh ^2\!\! \chi \left( \left(\frac{d\theta}{d\sigma^0}\right)^2\!\! +
\sin^2\!\! \theta \left(\frac{d\omega}{d\sigma^0}\right)^2\right)\right]\right\} .
\end{equation}
So, the picture is that the string-probe was replaced by a
``world sheet instanton'' mapping the Euclidean world sheet onto the
internal target space torus (with coordinates $y$ and $z$) and being
localized in space-time. With changing Eucledian world sheet time
$\sigma^0$ the locus of the ``world sheet instanton'' in the
four dimensional space-time will move along geodesics with respect
to
\begin{equation}
ds^2 = e^{2\varphi} \left( -d\tau^2 + \tau^2 d\Omega_{3,-1} ^2 \right) .
\end{equation}
That is, when we probe the four dimensional space-time with
those ``world sheet instantons'' we will not encounter problems
due to imaginary electric and magnetic charges, (for magnetic charges
one has to repeat the consideration for a heterotic dual string-probe
leading to the same result with $e^\varphi$ replaced by $e^{-\lambda}$).
Finally, we mention that it might be interesting to include the
non-Abelian background into the probe-action. For that one needs 
to derive a sigma model for the
heterotic $K3$-compactified string with unbroken gauge group $SU(2)$.
Theoretically one could do this by $K3$-compactifying the sigma model of
the ten dimensional heterotic string \cite{sen} in a
suitable instanton background. In praxis however, we do
not know how to do this and leave this question for future research.

\section{The 4-d cosmology}
In this section we discuss the 4-d cosmological models obtained by
rotating the 1-brane in the way described above.
We will first recall the case for zero Yang-Mills-fields, i.e.~the
cosmologies related to the dyonic string solutions of the Neveu-Schwarz sector 
of string theories.
Here we distinguish an extremal BPS solution and its
non-extremal generalization. We will then discuss the model obtained from
the six dimensional gauge dyonic string, where the charges are due to a
single $SU(2)$-instanton Yang-Mills field in the transverse space.

\subsection{The cosmology from the non-extremal dyonic string}
\begin{figure}[htb]
\begin{minipage}{8cm}
\centerline{2a}
\psfig{figure=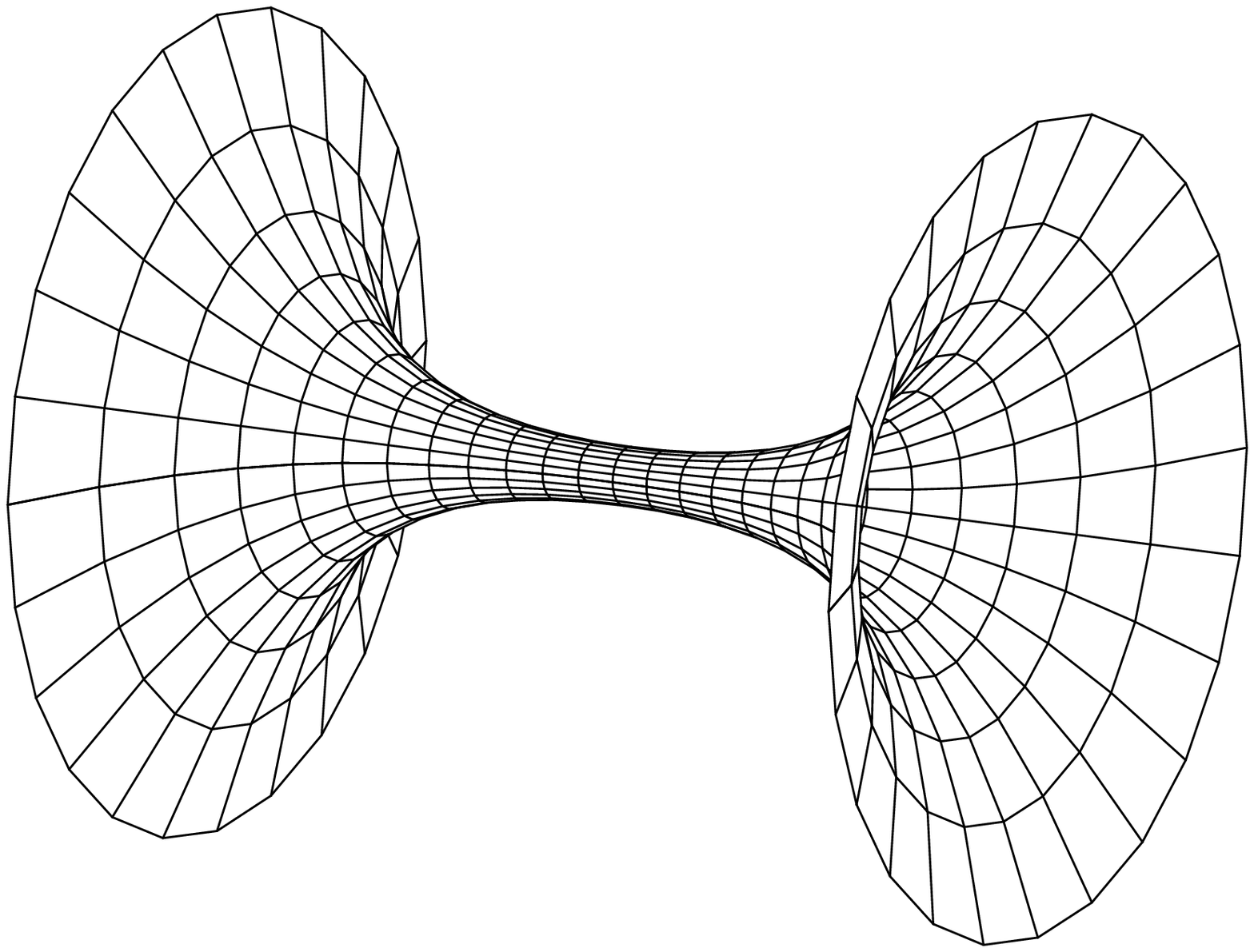,width=8cm,angle=270}
\end{minipage}\hfill
\begin{minipage}{8cm}
\centerline{2b}
\psfig{figure=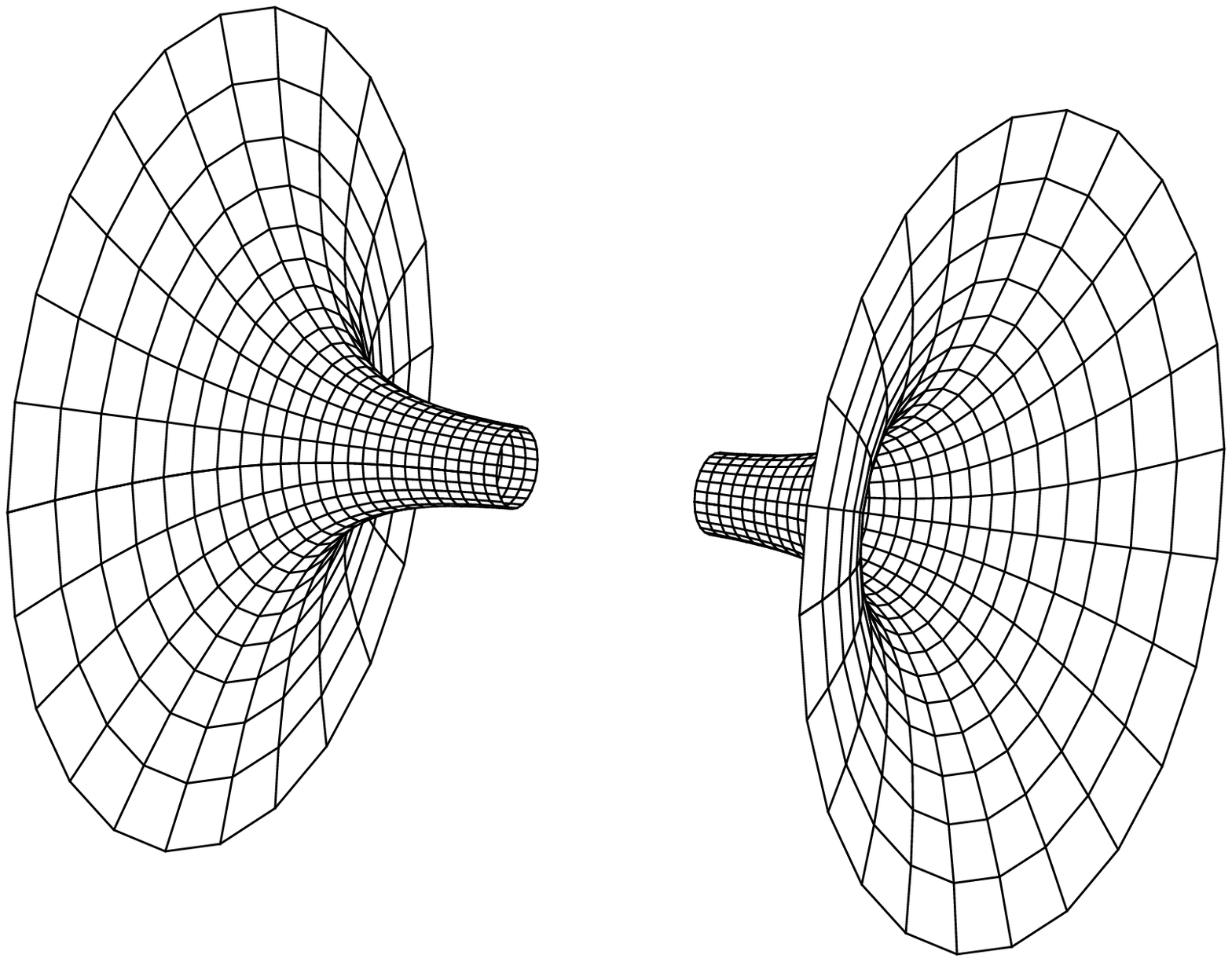,width=8cm,angle=270}
\end{minipage}\\
\begin{minipage}{8cm}
\centerline{2c}
\psfig{figure=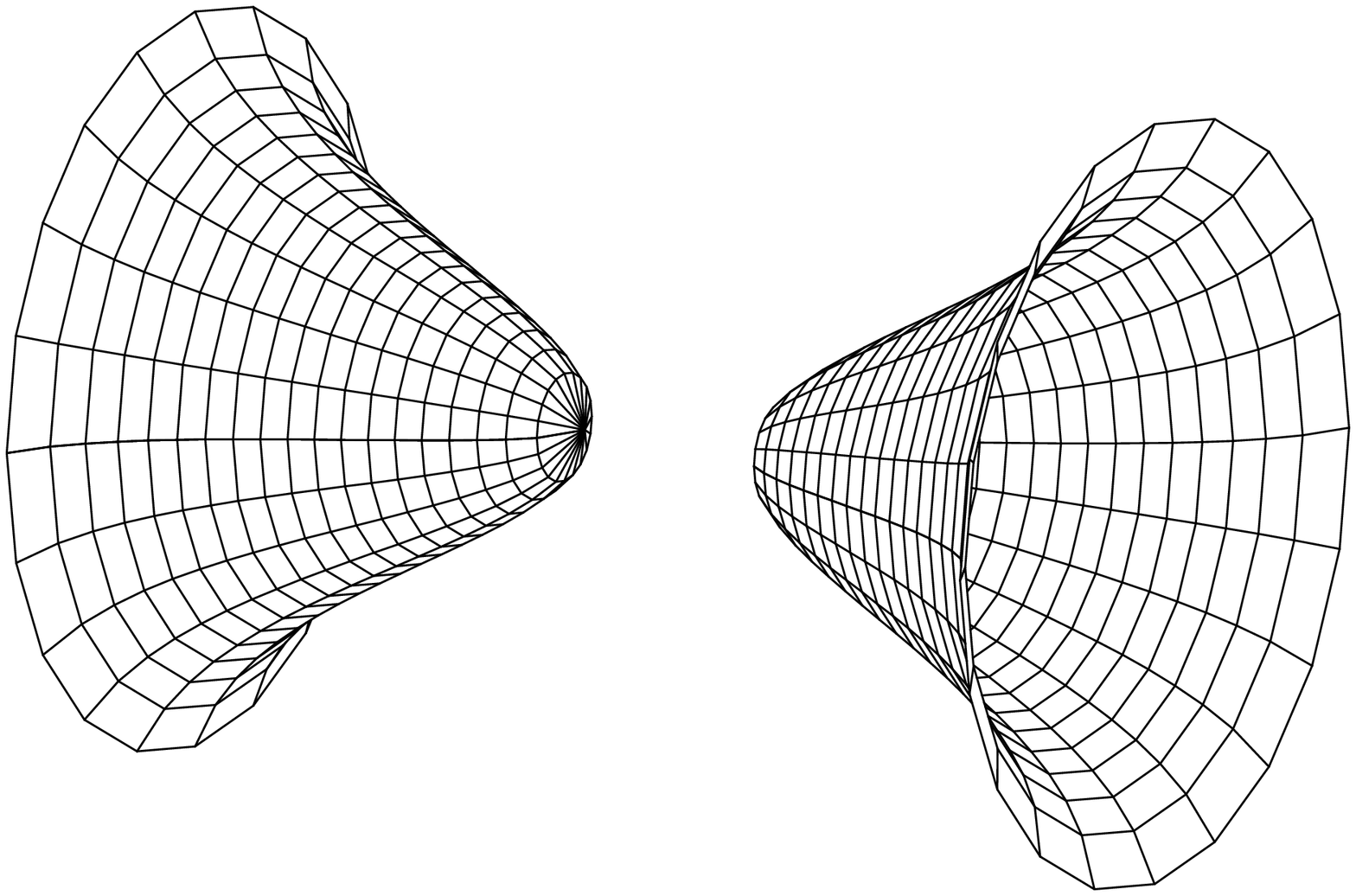,width=8cm,angle=270}
\end{minipage}\hfill
\begin{minipage}{8cm}
\end{minipage}
\caption{\label{cosmo}The cosmologies obtained from the nonextremal dyonic 
string (2a), its extremal BPS version (2b) and the BPS gauge dyonic string 
with a single $SU(2)$-instanton (2c).}
\end{figure}

Performing the rotation and the compactification within the non-extremal 
generalization of the BPS dyonic string \cite{dulupo} and changing to the
string frame, we obtain the four dimensional solution  
\begin{eqnarray}\label{4dnonextrem}
ds^2&=&\left(1+\frac{\mu^2}{\tau^2}\cosh^2\alpha\right)
       \left(-\frac{d\tau^2}{1+\frac{\mu^2}{\tau^2}} +
       \tau^2 d\Omega^2_{-1}\right),                      \nonumber\\
e^{2\varphi} &=&  1+\frac{\mu^2}{\tau^2}\cosh^2\alpha,    \\
e^{-2\lambda} &=& 1+\frac{\mu^2}{\tau^2}\cosh^2\beta.     \nonumber
\end{eqnarray}
To discuss qualitatively the evolution of this open FRW-universe we transform 
the time coordinate by $\tau=e^\eta - \frac{\mu^2}{4}e^{-\eta}$,
giving the metric
\begin{equation}
ds^2=\left[\left(e^\eta - \frac{\mu^2}{4}e^{-\eta}\right)^2 + 
     \mu^2\cosh^2\alpha\right]
      \left(-d\eta^2 + d\Omega^2_{-1}\right).
\end{equation}
For $\eta\to\pm\infty$, the geometry approaches flat Minkowski space, while
for finite $\eta$ the geometry is $R\times S^3_{-1}$. This metric therefore
describes two flat Minkowski spaces connected by a whormhole 
(figure \ref{cosmo}a). The space-like part of the whormhole has its smallest 
extension at $\eta_0=\ln\frac{\mu}{2}$, but is always nonzero. 
Let us determine whether matter can pass the wormhole in finite time.
For massive particles at rest the corresponding proper time would be given by
\begin{equation} \label{wormhole}
s=\int\limits_{\eta_0-\eta'}^{\eta_0+\eta'}d\eta 
      \left[\left(e^\eta - \frac{\mu^2}{4}e^{-\eta}\right)^2 + 
      \mu^2\cosh^2\alpha\right]^{1/2}.
\end{equation}
Since the integrand is regular, the integral is finite for finite $\eta'$. 
Thus the ``big bang'' at $\eta_0$ is only finitely far in
the past or future, as seen from any point inside the wormhole.
The dilaton and axion however become singular at $\eta_0$, which
corresponds to $\tau=0$.

\subsection{The cosmology from the extremal dyonic string}
The 6-d solution is the BPS dyonic solution of \cite{dfkr}. After performing the
rotation and compactifying to four dimensions we obtain in the string frame
\begin{eqnarray}
ds^2 &=& \left(1+\frac{P}{\tau^2}\right)
         \left(-d\tau^2 + \tau^2 d\Omega^2_{-1}\right),\nonumber\\
e^{2\varphi} &=& 1+\frac{P}{\tau^2}, \\
e^{-2\lambda} &=& 1+\frac{Q}{\tau^2}, \nonumber
\end{eqnarray}
with $Q$ the electric charge and $P$ the magnetic charge of the string.
The solution corresponds to the extremal limit of (\ref{4dnonextrem}), 
$\mu\to 0,\alpha, \beta\to\infty$ and
$\mu^2\cosh^2\alpha\to P, \mu^2\cosh^2\beta\to Q$.
The corresponding transformation of the time coordinate is now
$\tau=e^{\pm\eta}$, where the minus sign corresponds to the pre-big-bang 
phase,
and we obtain for the extremal case
\begin{equation}\label{4dextremevol}
ds^2=\left(e^{\pm 2\eta} + P\right)
\left(-d\eta^2 + d\Omega^2_{-1}\right).
\end{equation}
Considering the ``$+$'' branch, the geometry approaches again flat 
Minkowski space 
for $\eta\to\infty$, while for finite $\eta$ the geometry is $R\times S^3_{-1}$. 
The ``big bang'' however is now moved to $\eta=-\infty$ in 
the ``$+$'' branch and
$\eta=\infty$ in the ``$-$'' branch, i.e.~it is infinitely far in the past or 
future from each point in the throat.
We can therefore think of performing the extremal limit as making the throat 
connecting the two asymptotic Minkowski spaces 
infinitely long and thereby physically separating the pre-big-bang from the 
post-big-bang phase.
As in the non-extremal case, the dilaton and axion field diverge in the
limit $\tau \to 0$, which corresponds to the big bang in both branches.

\subsection{The cosmology from the gauge dyonic string with
$SU(2)$-in\-stan\-ton}
We now turn to the cosmology obtained from the gauge dyonic string of 
\cite{dlp},
where the charges of the string are produced by a single $SU(2)$-instanton.
After performing the Wick rotation and compactifying to four dimensions we
obtain the solution
\begin{eqnarray}
ds^2&=&\left(e^{2\varphi_0}+
        2\alpha' v\frac{2\rho^2+\tau^2}{(\rho^2+\tau^2)^2}\right)
        \left(-d\tau^2 + \tau^2 d\Omega^2_{-1}\right), \nonumber\\
e^{2\varphi} &=& e^{2\varphi_0}+ 2 \alpha' v
                \frac{2\rho^2+\tau^2}{(\rho^2+\tau^2)^2},  \label{cap}\\
e^{-2\lambda} &=& e^{-2\varphi_0}+ 2 \alpha' \tilde{v}
                \frac{2\rho^2+\tau^2}{(\rho^2+\tau^2)^2}. \nonumber
\end{eqnarray}
Again we transform the time coordinate, $\tau=e^{\pm\eta}$, which gives the 
metric
\begin{equation}
ds^2= e^{\pm 2\eta} \left( e^{2\varphi_0}+
      2\alpha' v\frac{2\rho^2+e^{\pm 2\eta}}{(\rho^2+e^{\pm 2\eta})^2}\right)
      \left(-d\eta^2 + d\Omega^2_{-1}\right).
\end{equation}
Taking the limit $\eta\to\pm\infty$ in the ``$+$'' and ``$-$'' 
branch, respectively, we
find again a decompactification of the $S^3_{-1}$ and therefore have 
asymptotic  Minkowski spaces. However, 
the radius of the $S^3_{-1}$ shrinks to zero for 
$\eta\to\mp\infty$, indicating again a separation of the two branches 
(figure \ref{cosmo}c).
We can get an upper bound for the proper time needed to reach the
collapsing point from an arbitrary point 
\begin{equation}
s=\int\limits_{-\infty}^{\eta'} ds \leq
e^{\eta^\prime}\sqrt{e^{2\varphi_0}+\frac{2\alpha' v}{\rho^2}},
\end{equation}
which is finite for finite instanton size $\rho$.
But at this point there is no singularity, all metric components
as well as scalar fields stay finite here. We have
a completely non-singular cosmological model. In the cases
discussed before, the (string) metric was also finite, but
the scalar fields were singular at $\tau=0$. 
At this point the internal torus collapsed and the string coupling
(dilaton) diverged.

\subsection{The brane picture}
\begin{figure}[t]
\begin{center}
\begin{minipage}{8cm}
\psfig{figure=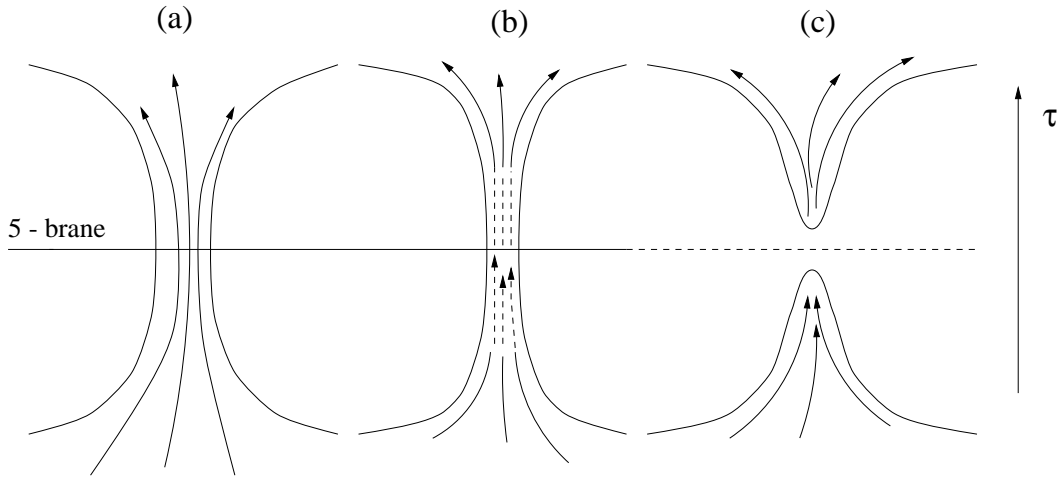,width=14cm,angle=270}
\end{minipage}\\
\end{center}
\vspace{-6mm}
\caption{
For the non-extreme 5-brane (a), matter can flow from one
asymptotic flat region to the other. In the extreme limit (b),
matter needs infinite time to reach or depart from the middle of the
throat. So, both universes are causally disconnected. Taking
an instanton source (c), results in a smooth ``closure'' of both
regions.
}
\end{figure}

\noindent
A main part of our cosmological model is the 5-brane.  If we neglect
for a moment the Yang Mills field, this 5-brane can be seen as a
connection between two asymptotic flat regions (universes).  But the
second flat region is not on the other side of the 5-brane, instead
one has to ``go through the 5-brane'' in order to reach it. This however
is only possible for a non-extreme 5-brane, where matter needs a
finite time to reach the other side, see eq.\ (\ref{wormhole}). Although the
geometry is smooth (in the string-frame), the scalar fields and
therefore also the gravitational coupling become singular in the
``moment one touches the 5-brane''.  This point one could call the
initial singularity at $\tau =0$. Since in our cosmological setting,
we have rotated the world volume of the 5-brane in a way that the time
is the transversal to the brane, matter can flow only in one direction.
From the thermodynamical point of view this means, the configuration is 
not in an equilibrium as one would expect for a non-extremal solution.

\medskip

\noindent
Next, going to the extremal case means, that the length of the throat
becomes infinite. In this case matter takes infinite time to reach or
to depart from the 5-brane in the middle, see eq.\ (\ref{4dextremevol}).  Thus, both
asymptotic regions are causally disconnected, it is impossible to go
``behind'' the 5-brane.

\medskip

\noindent
Finally, turning on the Yang-Mills field, this hole in space time is
closed by an instanton.  In this case all fields behave smooth
everywhere and the instanton can be seen as a cap that smoothly closes
the 5-brane throat. It takes a finite time to reach the bottom.  The
scalar fields and also the gravitational coupling are bounded by the
instanton size $\rho$, see eq.\ (\ref{cap}).  Hence, the instanton keeps
matter away from the original 5-brane and the two asymptotic regions
are not only causally disconnected but also geometrically.

\medskip

\noindent
In this picture the fundamental string, that lies on the 5-brane seems
to play no role. Really, turning on and off the electric charge has no
influence on the 4-d geometry and neither on the 4-d
dilaton. The world-volume of the fundamental string is mapped
completely on the internal torus and therefore in 4-d we do not have
anymore an electric charge, instead we can see it as topological
charge that tells us how many times we wrapped the world volume. In the
moment that one touches the 5-brane this internal torus shrinks to
zero size indicating the singularity at this point. Again, turning
on the YM instanton regularizes also this torus.
\section{Conclusions}
In the present paper we studied solutions of heterotic string theory
corresponding to four dimensional cosmological solutions. The ten 
dimensional configuration is a fundamental string within a 
solitonic five brane. $K3$ compactifying that background to six dimensions
and wrapping the five brane around $K3$ results in a dyonic string
solution in six dimensions. As sources for those strings we 
considered delta function like sources and instantons. In the
case of a delta function source extreme and non-extreme solutions
are known \cite{dulupo, dfkr} whereas the background with an instanton
source is known only in the extreme case \cite{dlp}. We rotated the
world volume such that the time is orthogonal to it and compactified
the world volume on $T^2$ ending up with a four dimensional
cosmological solution.

\medskip

In the non-extreme solution with delta function source, (which now acts as
a big bang singularity), we obtained two asymptotically flat regions
connected by a wormhole. Matter can tunnel through the big bang singularity
and would face singular background fields at the big bang. 
In the extreme limit of that solution matter takes infinite time to reach
the singularity and the universe looks like a half throat with one
asymptotically flat region.
In the case where the instanton is the source of the six dimensional string
we observed an asymptotically flat region at one end of the universe and
a smoothly vanishing world radius at the other end. In the six dimensional
model the finite instanton size corresponds to a spatially extended source 
for the string.
After rotating the world volume of the string into space like directions
this gets translated into a big bang source during a finite time interval
resulting in a completely non singular model. It might be interesting to
investigate whether these or similar effects can help to address the graceful
exit problem of the pre-big-bang scenario [\ref{ven1}--\ref{venlast}].
\section*{Acknowledgments}
We would like to thank Thomas Mohaupt and Stefan Theisen
for useful discussions.
S.F.\ and S.S.\ benefited from discussions with
Debashis Ghoshal.
They also acknowledge the hospitality of Institut f\"ur Physik
at Humboldt University during a stay where part of the presented
results was obtained. K.B.\ thanks Stefan Theisen for an
invitation to Munich university.\\
The presented work is partly supported by the DFG through
SFB 375-95 and TMR programs ERBFMX-CT96-0045 and
ERBFMX-CT96-0090.
K.B. and S.S.\ are supported by DFG, Deutsche Forschungsgemeinschaft.
The work of S.F.\ is supported by GIF, German Israeli Foundation
for scientific research. 
\newpage


\begin{thebibliography}{15}
\bibitem{anton1} \label{first} I.\ Antoniadis, C.\ Bachas, J.\ Ellis and
D.V.\ Nanopoulus, {\it ``An expanding universe in string theory''}
Nucl.\ Phys.\ {\bf B328} (1989) 117.
\bibitem{tseytlin-co} A.A.\ Tseytlin, {\it ``Cosmological solutions
with dilaton and maximally symmetric space in
string theory''}, Int.\ J.\ Mod.\ Phys.\ {\bf D1} (1992) 223,
{\tt hep-th/9203033}.
\bibitem{v1} \label{ven1}
M.\ Gasperini and G.\ Veneziano, {\it ``Pre-Big Bang in
String Cosmology''}, Astropart.\ Phys.\ {\bf 1} (1993) 317;
{\tt hep-th/9211021}.
{\it ``Inflation, deflation
and frame independence in string cosmology''}, Mod.\ Phys. Lett.\
{\bf A8} (1993) 3701, {\tt hep-th/9309023};
{\it ``Dilaton production
in string theory''}, Phys.\ Rev.\ {\bf D50} (1994) 2519, {\tt gr-qc/9403031}.
\bibitem{brust-ven} R.\ Brustein and G.\ Veneziano,
{\it ``The graceful exit problem in string cosmology''},
Phys.\ Lett.\ {\bf B329} (1994) 429. {\tt hep-th/9403060}.
\bibitem{brust} R.\ Brustein, M.\ Gasperini, M.\ Giovannini, V.F.\ 
Mukhanov and G. Veneziano, {\it ``Metric pertubations in dilaton
driven inflation''}, Phys.\ Rev. {\bf D51} (1995) 6744,
{\tt hep-th/9501066}; G. Veneziano, {\it ``Inhomogenous Pre-Big Bang
String Cosmology''}, {\tt hep-th/9703150}.
\bibitem{venlast} \label{venlast} M. Gasperini,
{\it ``Relic Dilatons in string cosmology''},  in ''Proc. of the 
12th Italian Conference on Gen. Rel. and Gravitational
Physics" (Rome, September 1996), ed. by M. Bassan et al. 
(World Scientific, Singapore), {\tt gr-qc/9611059}. 
\bibitem{nap-wit} C.R.\ Nappi and E.\ Witten, 
{\it ``A closed, expanding universe in string theory''}, Phys.\ Lett.\
{\bf B293} (1992) 309, {\tt hep-th/9206078}.
\bibitem{giveon-pasqu}A.\ Giveon and A.\ Pasquinucci,
 {\it ``On cosmological string backgrounds
with toroidal isometries''}, Phys.\ Lett.\ {\bf B294} (1992) 162, 
{\tt hep-th/9208076}.
\bibitem{lust} D.\ L\"ust, {\it ``Cosmological string backgrounds''},
 Presented at 4th Hellenic School
on Elementary Particle Physics, Corfu, Greece, 2-20 Sep 1992,
{\tt hep-th/9303175}.
\bibitem{behrndt} K.\ Behrndt and S.\ F\"orste,
{\it ``Cosmological string solutions in four dimensions
from 5d black holes''}, Phys.\ Lett.\ {\bf B320} (1994) 253,
{\tt hep-th/9308131};
{\it ``String 
Kaluza-Klein cosmology''}, Nucl.\ Phys.\ {\bf B430} (1994) 441,
{\tt hep-th/9403179}.
\bibitem{wands} E.J.\ Copeland, A.\ Lahiri and D.\ Wands,
{\it ``Low-energy effective string cosmology''}, Phys.\ Rev.\
{\bf D50} (1994) 4868, {\tt hep-th/9406216};
{\it ``String
cosmology with a time dependent antisymmetric tensor''},
Phys.\ Rev. {\bf D51} (1995) 1569, {\tt hep-th/9410136}.
\bibitem{behrndt1} K.\ Behrndt and T.\ Burwick, {\it ``Towards
quantum cosmology without singularities''}, Phys.\ Rev.\ 
{\bf D50} (1995) 1295, {\tt hep-th/9407039}.
\bibitem{levin} J.\ Levin, {\it ``Inflation from extra dimensions''},
Phys.\ Lett.\ {\bf B343} (1995) 69, {\tt gr-qc/9411041}.
\bibitem{easther} 
R.\ Easther, K.\ Maeda and D.\ Wands,
{\it ``Tree level string cosmology''}, Phys.\ Rev.\ {\bf D53} (1996) 4247,
{\tt hep-th/9509074}.
\bibitem{easther1}
I.\ Antoniadis, J.\ Rizos and K.\ Tamvakis, {\it ``Singularity free cosmological 
solutions of the superstring effective action''}, 
Nucl.\ Phys.\ {\bf B415}(1994) 497, 
{\tt hep-th/9305025};
R.\ Easther and K.\ Maeda, {\it ``One loop
superstring cosmology and the nonsingular universe''},
Phys.\ Rev.\ {\bf D54} (1996) 7252, {\tt hep-th/9605173}.
\bibitem{kaloper} N.\ Kaloper, {\it ``Stringy Toda Cosmologies''},
Phys.\ Rev.\ {\bf D55} (1997) 3394, {\tt hep-th/9609087}.
\bibitem{lumpo} H.\ L\"u, S.\ Mukherji and C.N.\ Pope and K.W.\ Xu,
{\it ``Cosmological solutions in string theories''}, 
{\tt hep-th/9610107}.
\bibitem{schwager-poppe} R.\ Poppe and S.\ Schwager, {\it ``String
Kaluza-Klein cosmologies with RR fields''}, Phys.\ Lett.
{\bf B393} (1997) 51, {\tt hep-th/9610166}.
\bibitem{tseytlin-cosmo} A.A.\ Tseytlin, {\it ``On the structure of
composite black p-brane configurations and related black holes''},
{\tt hep-th/9611111}.
\bibitem{copeland} E.J.\ Copeland, R.\ Easther and D.\ Wands, 
{\it ``Vacuum fluctuations in axion-dilaton cosmologies''},
{\tt hep-th/9701082}.
\bibitem{ovrut}
A.\ Lukas, B.A.\ Ovrut and D.\ Waldram, {\it ``Cosmological
solutions in type II string theory''}, Phys.\ Lett.\ {\bf B393} (1997)
65, {\tt hep-th/9608195};
{\it ``String and M-theory cosmological solutions with
Ramond forms''}, {\tt hep-th/9610238};
{\it ``Stabilizing dilaton and moduli vacua in string
and M theory''}, {\tt hep-th/9611204}.
\bibitem{larsen} F.\ Larsen and F.\ Wilczek, {\it ``Resolution
of cosmological singularities''}, {\tt hep-th/9610252}.
\bibitem{lu}  H.\ L\"u, S.\ Mukherji and C.N. Pope, {\it ``From p-branes to
cosmology''}, {\tt hep-th/9612224}.
\bibitem{rama} S. Kalyana Rama, {\it ``Can string theory avoid
cosmological singularities?''}, {\tt hep-th/9701154}.
\bibitem{brustein} \label{last} R.\ Brustein and R.\ Madden,
{\it `` Graceful exit and energy conditions in string cosmology''},
{\tt hep-th/9702043}.
\bibitem{eric} E.\ Bergshoeff, M.\ de Roo, E. Eyras, B.\ Janssen and
J.P.\ van der Schaar, {\it "Mutliple intersections of $D$-branes
and $M$-branes"}, {\tt hep-th/9612095}.
\bibitem{sagnotti} A.\ Sagnotti, {\it ``A note on the Green-Schwarz 
mechanism in open string theory''}, Phys.\ Lett.\ {\bf B294} (1992),
{\tt hep-th/9210127}.
\bibitem{dmw} M.J.\ Duff, R.\ Minasian and E.\ Witten,
{\it ``Evidence for heterotic/heterotic duality''}, Nucl.\ Phys.\ {\bf B465}
(1996) 413, {\tt hep-th/9601036}.
\bibitem{berkooz} M.\ Berkooz, R.\ Leigh, J.\ Polchinski, J.\ Schwarz,
N.\ Seiberg and E.\ Witten, {\it ``Anomalies, Dualities and Topology of
$D=6$\ $N=1$ superstring vacua''}, Nucl.\ Phys.\ {\bf B475} (1996) 115,
{\tt hep-th/9605184}.
\bibitem{mv} D.R. Morrison and C. Vafa, {\it ``Compactifications of 
F-theory on Calabi-Yau threefolds 1 \& 2''}, Nucl.\ Phys.\ 
{\bf B473} (1996) 74, {\tt hep-th/9602114} \& Nucl.\ Phys.\ {\bf B476}
(1996) 437, {\tt hep-th/9603116}.
\bibitem{afiq} G.\ Aldazabal, A.\ Font, L.E. Ib\'a\~nez and F.\ Quevedo,
{\it ``Heterotic/heterotic duality in $D=6, D=4$''}, 
Phys.\ Lett.\ {\bf B380} (1996) 33, {\tt hep-th/9602097}.
\bibitem{dlp} M.J.\ Duff, H.\ L\"u and C.N.\ Pope, {\it ``Heterotic phase 
transitions
and singularities of the gauge dyonic string''}, 
Phys. Lett. {\bf B378} (1996) 101, {\tt hep-th/960303307}.
\bibitem{wb} J.\ Wess and J.\ Bagger, 
{\it ``Supersymmetry and Supergravity''},
Princeton University Press (1983).
\bibitem{gabriel} G.\ Lopes Cardoso, G.\ Curio and D.\ L\"ust,
{\it ``Perturbative couplings and modular forms in $N=2$ string models
with a Wilson line''}, {\tt hep-th/9608154}.
\bibitem{dulupo} M.J.\ Duff, H.\ L\"u and C.N.\ Pope,
{\it ``The black branes of M-theory''}, Phys.\ Lett. {\bf B382} (1996) 73,
{\tt hep-th/9604052}.
\bibitem{dfkr} M.J.\ Duff, S.\ Ferrara, R.R.\ Khuri and J.\ Rahmfeld,
{\it ``Supersymmetry and dual string solitons''}, Phys.\ Lett.\ {\bf B356}
(1995) 479, {\tt hep-th/9506057}.
\bibitem{cvetic} M.\ Cveti\v{c} and A.A. Tseytlin, {\it `` Nonextreme black
holes from nonextreme intersecting M-branes''}, Nucl.\ Phys.\ {\bf B478} 
(1996)
181, {\tt hep-th/9606033}.
\bibitem{douglas} M.\ Douglas, J.\ Polchinski and A.\ Strominger,
{\it ``Probing five-dimensional black holes with D-branes''},
{\tt hep-th/9703031}.
\bibitem{sen} A.\ Sen, {\it ``Equations of motion for
the Heterotic String Theory from the conformal Invariance of the
Sigma Model''}, Phys.\ Rev.\ Lett.\ {\bf 18} (1985) 1846;
{\it ``Heterotic string in arbitrary background
field''}, Phys.\ Rev.\ {\bf D32} (1985) 2102.
\end{thebibliography}
\end{document}